\documentclass[aps,superscriptaddress,showpacs,floatfix,amsmath,amssymb,twocolumn]{revtex4}

\usepackage{epsfig}
\usepackage{dcolumn}
\usepackage{bm}
\usepackage{color}

\begin{document}

\title{Experimental study of the two-body spin-orbit force}

\author{G.~Burgunder }
\affiliation{Grand Acc\'el\'erateur National d'Ions Lourds (GANIL),
CEA/DSM - CNRS/IN2P3, BP 55027, F-14076 Caen Cedex 5,
France}
\author{O.~Sorlin}
\affiliation{Grand Acc\'el\'erateur National d'Ions Lourds (GANIL),
CEA/DSM - CNRS/IN2P3, BP 55027, F-14076 Caen Cedex 5,
France}
\author{F.~Nowacki}
\affiliation{IPHC, CNRS/IN2P3, Universit\'e de Strasbourg, F-67037 Strasbourg, France}
\author{S. Giron}
\affiliation{Institut de Physique Nucl\'eaire , CNRS/IN2P3, Universit\'e Paris-Sud, F-91406 Orsay, France}
\author{F. Hammache}
\affiliation{Institut de Physique Nucl\'eaire , CNRS/IN2P3, Universit\'e Paris-Sud, F-91406 Orsay, France}
\author{M.~Moukaddam}
\affiliation{IPHC, CNRS/IN2P3, Universit\'e de Strasbourg, F-67037 Strasbourg, France}
\author{N.~de S\'er\'eville}
\affiliation{Institut de Physique Nucl\'eaire , CNRS/IN2P3, Universit\'e Paris-Sud, F-91406 Orsay, France}
\author{D.~Beaumel}
\affiliation{Institut de Physique Nucl\'eaire , CNRS/IN2P3, Universit\'e Paris-Sud, F-91406 Orsay, France}
\author{L.~C\`aceres}
\affiliation{Grand Acc\'el\'erateur National d'Ions Lourds (GANIL),
CEA/DSM - CNRS/IN2P3, BP 55027, F-14076 Caen Cedex 5,
France}
\author{E.~Cl\'ement}
\affiliation{Grand Acc\'el\'erateur National d'Ions Lourds (GANIL),
CEA/DSM - CNRS/IN2P3, BP 55027, F-14076 Caen Cedex 5,
France}
\author{G.~Duch\^ene}
\affiliation{IPHC, CNRS/IN2P3, Universit\'e de Strasbourg, F-67037 Strasbourg, France}
\author{J.P.~Ebran}
\affiliation{CEA, DAM, DIF, Bruy\`eres-le-Ch\^atel, F-91297 Arpajon Cedex, France}
\author{B.~Fernandez-Dominguez}
\affiliation{Grand Acc\'el\'erateur National d'Ions Lourds (GANIL),
CEA/DSM - CNRS/IN2P3, BP 55027, F-14076 Caen Cedex 5,
France}
\affiliation{Department of Particle Physics, Universidade de Santiago de Compostela, E-15782, Santiago de Compostela, Spain}
\author{F.~Flavigny}
\affiliation{CEA Saclay, IRFU, Service de Physique Nucl\'eaire, F-91191 Gif-sur-Yvette, France}
\author{S. Franchoo}
\affiliation{Institut de Physique Nucl\'eaire , CNRS/IN2P3, Universit\'e Paris-Sud, F-91406 Orsay, France}
\author{J.~Gibelin}
\affiliation{LPC Caen, ENSICAEN, Universit\'e de Caen, CNRS/IN2P3, F14050 CAEN Cedex, France}
\author{A.~Gillibert}
\affiliation{CEA Saclay, IRFU, Service de Physique Nucl\'eaire, F-91191 Gif-sur-Yvette, France}
\author{S.~Gr\'evy}
\affiliation{Grand Acc\'el\'erateur National d'Ions Lourds (GANIL),
CEA/DSM - CNRS/IN2P3, BP 55027, F-14076 Caen Cedex 5,
France}
\affiliation{Centre dÕ \'Etudes Nucl\'eaires de Bordeaux Gradignan - Universit\'e Bordeaux 1 - UMR 5797 CNRS/IN2P3,
Chemin du Solarium, BP 120, 33175 Gradignan, France}
\author{J. Guillot}
\affiliation{Institut de Physique Nucl\'eaire , CNRS/IN2P3, Universit\'e Paris-Sud, F-91406 Orsay, France}
\author{V.~Lapoux}
\affiliation{CEA Saclay, IRFU, Service de Physique Nucl\'eaire, F-91191 Gif-sur-Yvette, France}
\author {A.~Lepailleur}
\affiliation{Grand Acc\'el\'erateur National d'Ions Lourds (GANIL), 
CEA/DSM - CNRS/IN2P3, BP 55027, F-14076 Caen Cedex 5,
France}
\author{I. Matea}
\affiliation{Institut de Physique Nucl\'eaire , CNRS/IN2P3, Universit\'e Paris-Sud, F-91406 Orsay, France}
\author{A. Matta}
\affiliation{Institut de Physique Nucl\'eaire , CNRS/IN2P3, Universit\'e Paris-Sud, F-91406 Orsay, France}
\affiliation{Department of Physics, University of Surrey, Guildford, GU2 5XH, United Kingdom}
\author{L.~Nalpas}
\affiliation{CEA Saclay, IRFU, Service de Physique Nucl\'eaire, F-91191 Gif-sur-Yvette, France}
\author{A. Obertelli}
\affiliation{CEA Saclay, IRFU, Service de Physique Nucl\'eaire, F-91191 Gif-sur-Yvette, France}
\author{T. Otsuka}
\affiliation{Department of Physics and Center for Nuclear Study, University of Tokyo, Hongo, Bunkyo-ku, Tokyo 113-0033, Japan}
\author{J.~Pancin}
\affiliation{Grand Acc\'el\'erateur National d'Ions Lourds (GANIL),
CEA/DSM - CNRS/IN2P3, BP 55027, F-14076 Caen Cedex 5,
France}
\author{A. Poves}
\affiliation{Departamento de Fisica Teorica e IFT-UAM/CSIC, Universidad Autonoma de Madrid, E-28049 Madrid, Spain}
\author{R. Raabe}
\affiliation{Grand Acc\'el\'erateur National d'Ions Lourds (GANIL),
CEA/DSM - CNRS/IN2P3, BP 55027, F-14076 Caen Cedex 5,
France}
\affiliation{Instituut voor Kern- en Stralingsfysica, KU Leuven, Leuven, Belgium}
\author{J. A. Scarpaci}
\affiliation{CSNSM-IN2P3-CNRS, Universit\'e Paris-Sud, 91405 Orsay, France}
\author{I. Stefan}
\affiliation{Institut de Physique Nucl\'eaire , CNRS/IN2P3, Universit\'e Paris-Sud, F-91406 Orsay, France}
\author{C.~Stodel}
\affiliation{Grand Acc\'el\'erateur National d'Ions Lourds (GANIL),
CEA/DSM - CNRS/IN2P3, BP 55027, F-14076 Caen Cedex 5,
France}
\author{T. Suzuki}
\affiliation{Department of Physics, Nihon University, Sakurajosui, Setagaya-ku, Tokyo 156-8550, Japan}
\author{J.C.~Thomas}
\affiliation{Grand Acc\'el\'erateur National d'Ions Lourds (GANIL),
CEA/DSM - CNRS/IN2P3, BP 55027, F-14076 Caen Cedex 5,
France}

\begin{abstract}

Energies and spectroscopic factors of the first $7/2^-$, $3/2^-$, $1/2^-$ and $5/2^-$ states in the $^{35}$Si$_{21}$ nucleus were determined by means of the (d,p) transfer reaction in inverse kinematics at GANIL using the MUST2 and EXOGAM detectors. 
By comparing the spectroscopic information on the $^{35}$Si and $^{37}$S isotones, a reduction of the $p_{3/2} - p_{1/2}$ spin-orbit splitting by about 25\% is proposed, while the $f_{7/2} -f_{5/2}$ spin-orbit splitting seems to remain constant. These features, derived after having unfolded nuclear correlations using shell model calculations, have been attributed to the properties of the 2-body spin-orbit interaction, the amplitude of which is derived for the first time in an atomic nucleus.  The present results, remarkably well reproduced by using several realistic nucleon-nucleon forces,  provide a unique touchstone for the modeling of the spin-orbit interaction in atomic nuclei. 

\end{abstract}

\date{\today}
\maketitle

\emph{Introduction.}- The spin-orbit (SO) interaction, which originates from the coupling of a particle spin with its orbital motion, plays essential roles in quantum physics. In atomic physics it causes shifts in electron energy levels due to the interaction between their spin and the magnetic field generated by their motion around the nucleus.  In the field of spintronics, spin-orbit effects for electrons in materials \cite{spintronics} are used for several remarkable technological applications. 
In atomic nuclei, the amplitude of the SO interaction is very large, typically of the order of the mean binding energy of a nucleon.  It is an intrinsic property of the nuclear force that must be taken into account for their quantitative description.
%while in $\Lambda$ hypernuclei \cite{Ajimura} the SO splittings are about 20 times weaker. 

An empirical one-body SO force was introduced in atomic nuclei in 1949 \cite{MGMHAXEL}  to account for the magic numbers and shell gaps that could not be explained otherwise at that time. In this framework each nucleon experiences  a coupling between its orbital momentum $\vec{\ell}$ and intrinsic spin $\vec{s}$. This  $\ell s$ coupling is attractive  for nucleons having their orbital angular momentum aligned with respect to their spin ($j _>=\ell+s$) and repulsive in case of anti-alignment ($j _<=\ell-s$). Shell gaps are created between the $j_>$ and $j_<$ orbits at nucleon numbers 6, 14, 28, 50, 82 and 126 for $\ell$=1-6, the size of which increases with the $\ell$ value. However, quoting ref. \cite{Eliot54}, this  parametrized $\ell s$  term "may not be a real force in the nucleus, but rather a caricature of a more complicated two-body force". Moreover  it does not account for modifications of shell gaps observed throughout the chart of nuclides \cite{PPNP} and has no connection with realistic bare two-body forces \cite{Epel}.

Bare forces can be cast into central, tensor and two-body spin-orbit parts, the latter two contributing to modifications of the SO splitting between nuclei. While the central force requires substantial and complex renormalizations to be applied in the atomic nucleus, it seems that the intensity of the tensor force can be derived from bare forces to account for some shell evolution in atomic nuclei \cite{Otsu05,Otsu10,Gaud06}. The two-body SO interaction is so far the most poorly constrained. The first attempt to derive its intensity was made by looking at the increase of the  2$p_{3/2}$-2p$_{1/2}$ splitting between the $^{47}$Ar and $^{49}$Ca nuclei \cite{Gaud06, Gaud07,Sorlin}. However, the effect of the two-body SO force was diluted and possibly contaminated by other effects.  The present work aims at studying the change in the neutron $2p_{3/2}-2p_{1/2}$ SO splittings between the $^{35}$Si and $^{37}$S nuclei caused by the filling of the proton $2s_{1/2}$ orbit.  Between these nuclei, changes in the SO splitting are likely totally carried by the two-body SO interactions as the two-body central and tensor contributions equate for each SO partner \cite{Otsu05,Smirnova}. Effects of the proximity of the continuum and of proton-to-neutron binding energies on the central part of the interaction were estimated to be of less than 5\% using mean field calculations constrained to experimental binding energies. The present study therefore provides a first and unique constraint of the two-body SO interaction in atomic nuclei, to be compared to the value derived from realistic nucleon-nucleon forces.
 
\emph{Experiment.- } The changes in 2$p$ and 1$f$ SO splitting between the $^{37}$S and $^{35}$Si nuclei have been studied using (d,p) transfer reactions in inverse kinematics with beams of $^{36}$S and $^{34}$Si. The $^{34}$Si nuclei were produced at the Grand Acc\'el\'erateur National d'Ions Lourds ({\sc GANIL}) in the fragmentation of a 55 $A \cdot $MeV $^{36}$S$^{16+}$ beam, of mean intensity 3~$\mu$A, in a 1075 $\mu$m-thick Be target. The {\sc LISE3} spectrometer~\cite{LISE} was used to select and transport the $^{34}$Si nuclei which were slowed down to 20.5 $A$MeV by using an achromatic Be degrader of 559.3 $\mu m$ between the two dipoles of the spectrometer. A rate of 1.1$\times$10$^5$ $^{34}$Si ions per second and a purity of 95\% were achieved. In a separate spectrometer setting, a beam of $^{36}$S was produced in similar conditions, at an energy of 19 $A$MeV and an intensity limited to 2$\times$10$^5$ pps. Nuclei were tracked event by event with a position resolution (FWHM) of 1~mm using a set of two position-sensitive Multi Wire Proportional Chambers (MWPC)~\cite{CATS} placed 0.92~m and 0.52~m upstream of the 2.6(1)~mg/cm$^{2}$ CD$_2$ target in which transfer reactions took place.

Nuclei were identified by means of their energy loss in an ionization chamber (IC), of 10$\times$10 cm$^{2}$ surface area, placed 40~cm downstream of the target. The energy-loss E$_{IC}$ of the ions was obtained from the peak-height value of the digitized signal. A 1.5 cm thick plastic scintillator, located behind the IC, additionally provided a high-resolution time signal used for precise time-of-flight (TOF) measurements, and allowed the monitoring of the beam intensity complementary to the MWPC detectors. By achieving selections in E$_{IC}$ and in TOF between the MWPCs and the plastic scintillator, the Si nuclei (in the case of $^{34}$Si(d,p)) were selected and the part corresponding to incomplete fusion reactions induced by the C nuclei of the CD$_2$ target was rejected. 

Energies and angles of the protons arising from the (d,p) reactions were measured using four modules of the MUST2 detector array~\cite{MUST2} consisting each of a highly segmented (128 $\times$ 128) double-sided 300$\mu m$-thick Si detector, followed by a 16 fold segmented Si(Li) detector of 4.5~mm thickness. These detectors were placed at 10~cm from the CD$_2$ target, covering polar angles ranging from 105$^\circ$ to 150$^\circ$ with respect to the beam direction. In addition a 16 Si strips annular detector (external diameter  96~mm, central hole diameter of 48~mm and thickness 300$\mu$m) was placed  at a distance of 11.3~cm to cover polar angles from 157$^{\circ}$ to 168$^{\circ}$ to detect the full energy of protons in the (d,p) reaction.  

Four segmented Ge detectors from the EXOGAM array~\cite{EXOGAM} were installed perpendicular to the beam axis at a mean distance of 5~cm to detect the $\gamma$-rays emitted in the decay of excited states. The center of these detectors was shifted 9~cm downstream from the target in order to avoid them shadowing part of the  MUST2 detectors, leading to  a $\gamma$-efficiency of $\epsilon_\gamma$=3.8(2)\% at 1~MeV.

\begin{figure}
\begin{center}
\centering \epsfig{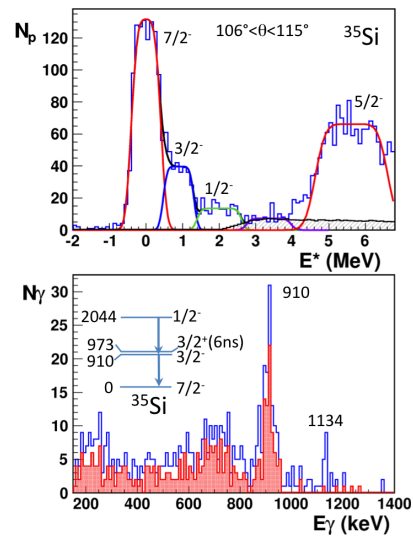}
\caption{(Color on-line) \textbf{Top} Excitation energy spectrum E* of $^{35}$Si obtained with the detection of protons in the angular domain 106-115$^{\circ}$ in the laboratory. The fitting procedure of the peaks, as well as the origin of the asymmetric black curve are described in the text.  \textbf{Bottom}: Doppler-corrected $\gamma$ energy spectrum gated on 0.2$<$E*$<$2.6~MeV (blue)  and on 0.2$<$E*$<$1.4~MeV (filled area in red). The 1.134 keV $\gamma$-ray appears only when gating on E*$>$ 1.4 MeV, indicating that it comes from a level at 2044 keV.}
\label{34Sidp}
\end{center}
\end{figure}

\emph{Results.- } Excitation energy spectra (E*) corresponding to the $^{34}$Si(d,p)$^{35}$Si reaction (Fig.~\ref{34Sidp}) were constructed using the energy and angle of the emitted protons  in coincidence with the Si nuclei. Three structures are seen \emph{below} the neutron emission threshold S$_n$=2.47(4)~MeV at E*=0(25), 906(32) and 2060(50)~keV. Other structures are present above $S_n$; tentatively at 3330(120) keV and more prominently at $\simeq$ 5500 keV. The presently fitted shape of these peaks is a convolution between a rectangular step function, that takes into account the energy loss of the beam in the target before the reaction point, and a Gaussian. The energy-dependent widths of all fitted peaks are in very good accordance with Monte Carlo simulations \cite{Jo}. A more accurate energy determination of the bound levels populated in $^{35}$Si is provided by the $\gamma$-energy spectrum, gated by protons associated to different E* ranges. When applying suitable Doppler corrections to the $\gamma$'s emitted in flight and detected in the EXOGAM array, two peaks are clearly observed at 910(3)~keV and 1134(6)~keV in the bottom part of Fig.~\ref{34Sidp}. The energy of the first $\gamma$-peak matches that of  E*=906(32)~keV of Fig.~\ref{34Sidp}, as well as the energy of a $3/2^-$ state at 910.10(30)~keV fed indirectly in the $\beta$-decay study of $^{35}$Al~\cite{Nummela}.  
From the number of protons detected in the peak at 906(32) keV, N$_p$=1894(185), an expected number of photons at 910 keV of N$_\gamma$= 72(11) is derived, after having corrected from the $\epsilon_\gamma$ value.  The number of detected photons, 82(10), matches this expected value of 72(11) within one $\sigma$ uncertainty. We deduce that a contamination of the excitation energy spectrum at E*=906(32) due to transfer to the 3/2$^+$ state at the nearby energy of 970 keV is less than 30\% of the 3/2$^-$ component, with a confidence limit of 3 $\sigma$. 
With a half-life of 6 ns, the $\gamma$-decay of the 3/2$^+$ isomer would occur after the target location, mostly out of the range of the EXOGAM detectors. 
The energy of the second $\gamma$-peak is in accordance with the one observed in~\cite{Gelin} at 1133(5)~keV. The summed energy of the two $\gamma$ peaks, 910(3)+1134(6)=2044(7)~keV, matches the energy of the third peak at E*=2060(50)~keV in Fig.~\ref{34Sidp}, hereby establishing a level at 2044(7)~keV which decays by a cascade of two $\gamma$-rays.

\begin{figure}
\centering \epsfig{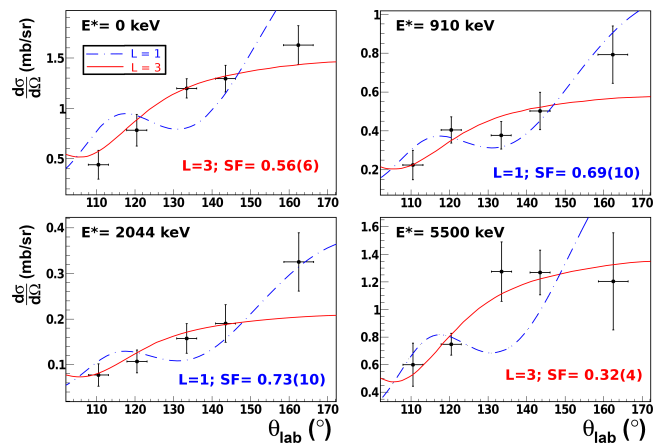} 
\caption{Proton angular
distributions of the states at E*=0, 910, 2044 and 5500~keV in $^{35}$Si. The curves
correspond to ADWA calculations assuming transfer to $\ell=1$ (dashed dotted) or $\ell=3$ (full line) states.}
\label{34Sidp2}
\end{figure}

\indent Proton angular distributions corresponding to transfer reactions populating the four states in $^{35}$Si are shown in Fig.~\ref{34Sidp2}. Adiabatic Distorted Wave Approximation (ADWA) calculations~\cite{JohnsonTandy} were performed using the code TWOFNR~\cite{TWOFNR} and the global optical
potentials of ~\cite{Wales} and~\cite{Varner} for the entrance and exit channels of  
 the (d,p) reaction, respectively.  A non-local correction~\cite{PereyBuck} has been used with Gaussian function of widths $\beta$=0.85~fm for the nucleons and 0.54 fm for the deuteron. These calculations were
fitted to the experimental angular distributions to infer the transferred
angular momentum $\ell$ and  Spectroscopic Factor (SF) of
individual orbitals in $^{35}$Si,  given with their uncertainties in Fig.~\ref{34Sidp2}. Additional uncertainties on the SF values (not given here) due to the use of other global potentials amount to about 15\% \cite{Gaud06}.  The \emph{same} set of optical potentials was used for the $^{35}$Si and $^{37}$S nuclei. With this set, we reproduce within one sigma the \emph{mean} $<$SF$>$ values in $^{37}$S derived from Refs.~\cite{Eckle,Thorn} for the 7/2$^-$ ground state ($<$SF$>$=0.73; our value 0.69(14)), the 3/2$^-_1$ state at 645~keV ($<$SF$>$=0.545; our value 0.53(10)) as well as the $1/2^-_1$ state at 2638~keV ($<$SF$>$=0.625; our value 0.68(13)) \cite{Jo}.  It has been pointed out in \cite{quenching} that observed SF are usually quenched, by a factor of about 0.5-0.7, as compared to the ones expected from single particle structure around closed shell nuclei. In the $^{37}$S nucleus, the SF values of the 7/2$^-$, 3/2$^-$ and 1/2$^-$ states exhaust this quenched SF sum rule, within the present experimental uncertainties.

\begin{figure}
\centering \epsfig{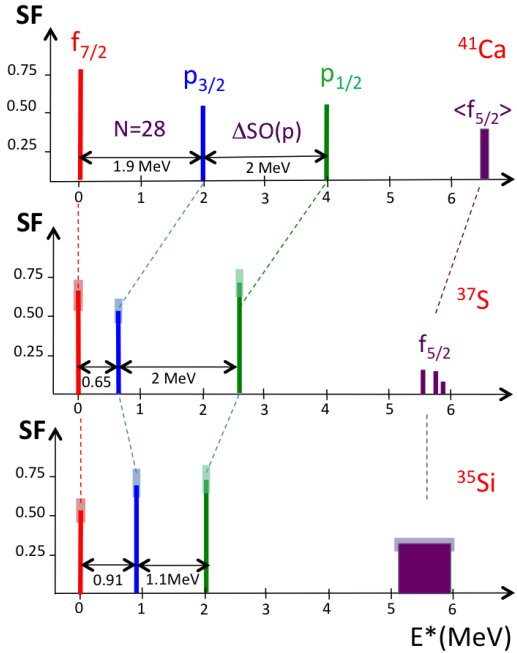} 
\caption{Distribution of the \textit{major} fragments of the single particle strength in  $^{41}$Ca (top),  $^{37}$S (middle) and in $^{35}$Si (bottom).  SF values in $^{41}$Ca are taken from \cite{Uozumi}. The centroid of the 5/2$^-$ strength, obtained from a summed SF strength of 0.32, is indicated as $<f_{5/2}>$. The SF of the 5/2$^-$ components in $^{37}$S are taken from \cite{Eckle}, while all others SF are derived from the present work with error bars due to statistics and fit distributions.}
\label{SF}
\end{figure}

From the shape of the proton angular distributions of Fig. \ref{34Sidp2}, the first peak in $^{35}$Si could be attributed to a $\ell=3$ transfer to the $f_{7/2}$ ground state with SF=0.56(6).  The angular distributions of the second and third peaks correspond to $\ell=1$, with SF values of 0.69(10) and 0.73(10), respectively. The third peak at 2044 keV is likely to be 1/2$^-$ as its large SF value discards another large $\ell=1$, 3/2$^-$ component. The SF values of these 7/2$^-$, 3/2$^-$ and 1/2$^-$ states in $^{35}$Si are compatible, within one $\sigma$, with the ones measured in $^{37}$S. However the excitation energy of the  1/2$^-$ state in $^{35}$Si (E*= 2044 keV) is significantly smaller than that in $^{37}$S (E*= 2638 keV). The structure above the neutron threshold at about 3330 keV likely corresponds to the elastic deuteron break-up process, the cross section of which was estimated to be 0.1mb/MeV \cite{Angela} and the shape of which was obtained from phase-space simulations (hatched zone below the black curve of the top part of Fig.~\ref{34Sidp}). The broad structure around 5.5~MeV in $^{35}$Si could be fitted with an angular distribution corresponding to a $\ell=3$ state  coming from a fraction of the $f_{5/2}$ strength. Using the prescription of Ref.~\cite{JohnsonSoppert} for the states lying in the continuum, a value of  SF=0.32(2) has been extracted. It has a similar amplitude as the $f_{5/2}$ component SF=0.36 found in three states centered around 5.6~MeV in $^{37}$S~\cite{Eckle}. 

\emph{Change in $p$-orbitals SO splitting? - }  To a first approximation the first states in $^{41}$Ca, $^{37}$S and $^{35}$Si can be viewed as one $1f_{7/2}$ or  $2p_{1/2,3/2}$ neutron on top of the core nuclei $^{40}$Ca, $^{36}$S and $^{34}$Si, respectively, as these N=20 nuclei can be considered as doubly magic nuclei. When taking the \emph{major} fragment of the $2p_{3/2}$ and $2p_{1/2}$  single-particle (SP) strengths, the 3/2$^-$  - 1/2$^-$ splitting remains close to 2~MeV in the $^{41}$Ca \cite{Uozumi} and $^{37}$S \cite{Eckle,Thorn} nuclei after the removal of 4 protons from the $1d_{3/2}$ orbit. As shown in Fig. \ref{SF}, it drops to 1.134~MeV in $^{35}$Si by removing 2 protons from the $2s_{1/2}$ orbit. 
This sudden reduction of the 3/2$^-$  - 1/2$^-$  splitting is attributed to the difference in the two-body proton-neutron monopole terms V$^{pn}_{2s_{1/2}2p_{1/2}}$ and V$^{pn}_{2s_{1/2}2p_{3/2}}$ involved between the $^{35}$Si and $^{37}$S nuclei as well as to the effects of correlations inherent to atomic nuclei. As there is no change in 3/2$^-$  - 1/2$^-$  splitting between the $^{41}$Ca and $^{37}$S nuclei, other monopole terms such as the ones involving the proton $1d_{3/2}$ orbit are negligible.

Shell model calculations have been used in the full $sd-pf$ shells \cite{Nowacki} (including cross-shell mixing between normal and intruder neutron configurations  \cite{Rotaru}) as a tool to determine the role of correlations and to deduce the change of the $p$ SO splitting $\Delta$ SO(p) between  the $^{37}$S and $^{35}$Si nuclei from experimental data. The V$^{pn}_{2s_{1/2}2p_{1/2}}$ and V$^{pn}_{2s_{1/2}2p_{3/2}}$ monopole terms have been constrained to match, after taking into account the correlations in the full valence space, the experimental energies of the \emph{major} fragments in the $^{37}$S and $^{35}$Si isotones, leading to -0.844 and -1.101 MeV, respectively. The calculated $2s_{1/2} $ occupancy varies from 1.66 in $^{37}$S (close to the experimental value of $\simeq$ 1.7~\cite{Khan85}) to 0.19 in $^{35}$Si, yielding $\Delta 2s_{1/2}$=1.47. Following the previous discussion, $\Delta$ SO(p) can be expressed as:

\begin{equation} \label{SOp3}
\Delta SO(p) \simeq \Delta 2s_{1/2}  (V_{2s_{1/2} 2p_{1/2}}^{pn} -
V_{2s_{1/2} 2p_{3/2}}^{pn}) 
\end{equation} 

Consistent values of $\Delta$ SO(p)=1.47$\times$257= 378 keV  and 380 keV are found using Eq.~\ref{SOp3} and the prescription of Baranger~\cite{corr}, respectively. The latter value is obtained from the  energies of the single-particle centroids of the $p_{3/2}$ and $p_{1/2}$ states derived from the calculated particle and hole energy weighted sum rules of all $3/2^-$ and $1/2^-$ states. The agreement between the two methods shows that the earlier assumption that the changes in the $p$ SO splitting are solely carried by the $V_{s p}$ monopoles is correct.
After applying a quenching factor of 0.7 to the SM calculations, we find that the calculated SF values of the major fragments 7/2$^-$ (SF=0.59) , 3/2$^-$ (0.59), 1/2$^-$ (0.61) and 5/2$^-$ (0.28) agree with the experimental values of 0.56(6), 0.69(10), 0.73(10), 0.32(3). 

 \emph{Realistic two-body SO interactions -}  
The M3Y interaction \cite{M3Y}, constructed as a model to realistic G-matrix interaction, was used to calculate
the 2-body SO parts of the monopole matrix elements for A$\simeq$40.  We find that  $\tilde{V}^{pn}_{2s_{1/2} 2p_{1/2}}$ ($\tilde{V}^{pn}_{2s_{1/2}2p_{3/2}}$) is repulsive (attractive) and amounts to +0.178 MeV (-0.089 MeV). Their difference, 0.267 MeV, is also in remarkable agreement with the value of 0.257 MeV derived from the experiment.  We then look at more modern interactions obtained from chiral effective field theory \cite{N3LO} as well as from the Kahana-Lee-Scott (KLS) potential \cite{KLS}, the latter being used for cross-shell matrix elements in the SDPF-U interaction \cite{Rotaru}.  The N3LO results $^{a)}$ of  Table~\ref{N3LO-monop} correspond to the $V_{lowk}$ renormalization with a cut-off $\Lambda= 1.8 fm^{-1}$ in an harmonic oscillator basis with $\hbar\omega$ = 11.5 MeV, appropriate for A $\sim$ 36.  We see a very small sensitivity to the cut-off renormalization of the interaction when many-body perturbation theory (MBPT) techniques from \cite{morten} are applied respectively in a 2 $^{b)}$ and 4 $^{c)}$ major shells basis.  The order of magnitude of the difference between the $V^{pn}_{2s_{1/2}1p_{3/2}}$ and $V^{pn}_{2s_{1/2}1p_{1/2}}$ ($\sim 300$ keV) monopoles derived from the bare interactions is similar to the value of 257 keV derived from the experiment. Their spin-tensor decomposition, using the same procedure as in \cite {Smirnova}, shows that their difference is totally carried by the two-body SO term (K=1). 
\begin{table} [t]
\begin{center}
\caption{Values of the proton-neutron monopole matrix elements in MeV between the $2s_{1/2}$ proton and $2p$ neutron orbitals for the KLS and N3LO interactions $^{a)}$ (bare), $^{b)}$  ($2\hbar\omega$), $^{c)}$  ($4\hbar\omega$).  Their spin-tensor decomposition \cite{Smirnova} into central (K=0) and spin-orbit (K=1) is also given. The tensor term (K=2) amounts to zero in all cases.
 \label{N3LO-monop}}
\begin{tabular}{c|ccc|ccc}
\hline
\multicolumn{1}{c}{Monopole} &  \multicolumn{3}{c}{$V_{2s_{1/2} 2p_{1/2}}^{pn}$} &  \multicolumn{3}{c}{$V_{2s_{1/2} 2p_{3/2}}^{pn}$}\\
\hline
decomposition&total&K=0&K=1&total&K=0&K=1\\
\hline
N3LO$^a$  & -1.124&-1.317&0.193&  -1.413  &-1.317& 0.193\\
N3LO$^b$  & -1.128&-1.312&0.184&  -1.404  &-1.312&-0.092\\
N3LO$^c$  & -1.201&-1.401&0.200&  -1.500  &-1.401&-0.100\\
KLS           & -1.180 &-1.374&0.194& -1.471&-1.374&-0.097   \\
%\hline\\
\end{tabular}
\end{center}
\end{table}

\emph{Conclusions.- } The energies and spectroscopic factors of the first 7/2$^{-}$, 3/2$^{-}$, 1/2$^{-}$ and 5/2$^{-}$ neutron states have been determined in the $^{37}$S  and $^{35}$Si isotones. A change by 25\% in the neutron SO splitting $p_{3/2} - p_{1/2}$ is derived between the $^{37}$S and $^{35}$Si nuclei from experimental data corrected for correlation effects, while no change in the $f_{7/2} - f_{5/2}$  SO splitting is observed within the present experimental limitations. This work presents the cleanest extraction of the 2-body SO interaction by choosing an experimental situation in which contributions from other components of the nuclear force are likely suppressed or modest. The derived strength of the 2-body SO interaction is remarkably well reproduced by realistic nucleon-nucleon forces such as N3LO and KLS, suggesting that these forces could be used more widely to predict its strength in other regions of the chart of the nuclides. The present results also carry important potentialities to test the density and isospin dependencies of the SO interaction in mean field theories.

\acknowledgments {\small The GANIL technical groups  are warmly acknowledged for their help during the preparation and running of the experiment.  A. Bonaccorso is greatly acknowledged for discussions and calculations related to the deuteron break-up component.}

\end{document}